\documentclass[10pt, a4paper, twocolumn]{article} %
\usepackage{hyperref}

\usepackage[english]{babel} %

\usepackage{microtype} %

\usepackage{amsmath,amsfonts,amsthm} %

\usepackage[svgnames]{xcolor} %

\usepackage[hang, small, labelfont=bf, up, textfont=it]{caption} %

\usepackage{booktabs} %

\usepackage{lastpage} %

\usepackage{graphicx} %

\usepackage{enumitem} %
\setlist{noitemsep} %

\usepackage{sectsty} %
\allsectionsfont{\usefont{OT1}{phv}{b}{n}} %

\usepackage{geometry} %

\usepackage[T1]{fontenc} %
\usepackage[utf8]{inputenc} %

\usepackage{XCharter} %

\usepackage{fancyhdr} %
\fancypagestyle{firstpage}{ %
	\fancyhf{}
}

\newcommand{\authorstyle}[1]{{\large\usefont{OT1}{phv}{b}{n}\color{DarkRed}#1}} %

\newcommand{\institution}[1]{{\footnotesize\usefont{OT1}{phv}{m}{sl}\color{Black}#1}} %

\usepackage{titling} %

\newcommand{\HorRule}{\color{DarkGoldenrod}\rule{\linewidth}{1pt}} %

\pretitle{
	\vspace{-30pt} %
	\HorRule\vspace{10pt} %
	\fontsize{25}{36}\usefont{OT1}{phv}{b}{n}\selectfont %
	\color{DarkRed} %
}

\posttitle{\par\vskip 15pt} %

\preauthor{} %

\postauthor{ %
	\vspace{10pt} %
	\par\HorRule %
	\vspace{20pt} %
}

\usepackage{subfigure}
\usepackage{lettrine} %
\usepackage{fix-cm}	%
\usepackage{chemformula}
\usepackage{amsfonts}       %
\usepackage{booktabs}
\usepackage{multirow}
\newcommand{\bu}{\mathbf{u}}
\newcommand{\bU}{\mathbf{U}}
\newcommand{\bz}{\mathbf{z}}
\newcommand{\bZ}{\mathbf{Z}}
\newcommand{\bv}{\mathbf{v}}
\newcommand{\bV}{\mathbf{V}}

\newcommand{\boldf}{\mathbf{f}}

\usepackage{xstring} %

\usepackage[backend=bibtex,style=authoryear,natbib=true]{biblatex} %

\usepackage[autostyle=true]{csquotes} %

\def\abstract{\normalfont%
    \small\textit{Abstract}---}
\title{Learning graph representations of biochemical networks and its application to enzymatic link prediction}
\author{
	\authorstyle{Julie Jiang\textsuperscript{1}, Li-Ping Liu \textsuperscript{1}, and Soha Hassoun\textsuperscript{1,2}} %
	\newline\newline %
	\textsuperscript{1}\institution{Department of Computer Science, Tufts University, Medford, 02155, USA}\\ %
	\textsuperscript{2}\institution{Department of Chemical and Biological Engineering, Tufts University, Medford, 02155, USA} %
}

\begin{document}

\maketitle %

\thispagestyle{firstpage} %

\begin{abstract}
{\small
The complete characterization of enzymatic activities between molecules remains incomplete, hindering biological engineering and limiting biological discovery. We develop in this work a technique, Enzymatic Link Prediction (ELP), for predicting the likelihood of an enzymatic transformation between two molecules. ELP models enzymatic reactions catalogued in the KEGG database as a graph. ELP is innovative over prior works in using graph embedding to learn molecular representations that capture not only molecular and enzymatic attributes but also graph connectivity. 

We explore both transductive (test nodes included in the training graph) and inductive (test nodes not part of the training graph) learning models. We show that ELP achieves high AUC when learning node embeddings using both graph connectivity and node attributes. Further, we show that graph embedding for predicting enzymatic links improves link prediction by 24\% over fingerprint-similarity-based approaches. To emphasize the importance of graph embedding in the context of biochemical networks, we illustrate how graph embedding can also guide visualization. 

The code and datasets are available through \href{https://github.com/HassounLab/ELP}{https://github.com/HassounLab/ELP}.
}
\end{abstract}

\section{Introduction}

Characterizing enzymes through sequencing, annotation, and homology has enabled the creation of  complex system models that have played a critical role in advancing many biomedical and bioengineering applications.  Insufficient characterization of enzymes, however, fundamentally limits our understanding of metabolism and creates knowledge gaps across many applications.  For example, while nearly 300  $\beta$-glucuronidases  (gut-bacterial enzymes that hydrolyze glucuronate-containing polysaccharides such as heparin and hyaluronate as well as small-molecule drug glucuronides) have been cataloged, functional information is available for only a small fraction (<10\%) \citep{pellock2019discovery}, thus limiting our ability to analyze host-microbiota interactions. Importantly, most enzymes if not all are promiscuous, 
acting on substrates other than the enzymes' natural substrates \citep{d1998underground,tawfik2010enzyme,hult2007enzyme}. At least one-third of protein superfamilies are functionally diverse, each superfamily catalyzing multiple reactions \citep{Almonacid2011}. Despite progress in assigning functional properties to enzyme sequences \citep{brown2014new,cuesta2015classification,merkl2016ancestral,Baier2016,Finn2016}, the complete characterization or curation of enzyme function and the reactions they catalyze remains elusive.

Computational prediction of enzymatic transformations promises to complement existing databases and provide new opportunities for biological discovery. The most common predictor of enzyme-compound interaction is compound and/or enzyme similarity to those within known enzymatic reactions. In biological engineering, molecular similarity between a query molecule and native substrates that are known to be catalyzed by the enzyme inform putative enzymatic transformation steps along a synthesis pathway \citep{Pertusi2014}.  A high similarity score indicates a likely transformation. Molecular fingerprints, one-dimensional vectors with entries representing the presence or absence of molecular feature such as combinations of atom properties, bond properties, are used to compute similarity. Enzyme sequence information such as binding site covalence and thermodynamic favorability were also used to inform the prediction \citep{Cho2010}. Similarity-based predictions are also utilized in predicting drug-protein interactions. A recent survey \citep{Kurgan2018} summarizes available drug-protein databases and 35 recent similarity-based prediction techniques. Several such techniques apply SIMCOMP, a heuristic algorithm for computing  structural molecular similarity based on subgraph isomorphism \citep{Hattori2010}. Other techniques use machine learning on molecular feature vectors or enzyme features to compute the likelihood of interactions (e.g., \citep{Cobanoglu2013,Tsubaki2018,Ozturk2018}). Numerous computational methods including quantum mechanics, molecular docking, and machine learning, have been used to predict atoms and bonds that undergo biochemical transformations, referred to as site of metabolism, due to Cytochrome P450 enzymes \citep{Tyzack2019}. Prediction of promiscuous products have utilized either hand-curated rules (e.g., \citep{Morreel2014,Li2004}), or rules culled from enzymatic reactions  (e.g., \citep{Adams2010, Yousofshahi2015}).

We present in this paper a novel technique, Enzymatic Link Prediction (ELP), for predicting enzymatic transformations between two molecules. ELP advances over the state-of-the art in several ways.  First, ELP maps known enzymatic reactions already catalogued in databases (the KEGG database for this work) to a graph structure, where compounds are represented as graph nodes while reactions are represented as graph edges.  While such graph representations have been exploited during pathway analysis and construction, e.g., \citep{Yousofshahi2015}, they have not been exploited when studying enzyme promiscuity. Second, ELP uses graph embeddings \citep{goyal2018graph,cai2018comprehensive} to learn molecular representations that reflect not only molecular structural properties but also relationships with other molecules in the network graph. Such embeddings have proven effective in predicting missing information, identifying spurious interactions,  predicting links appearing in future evolving network, and analyzing biomedical networks  \citep{goyal2018graph,cai2018comprehensive,Xiang2019}. Third, ELP first uses embedding propagation to compute embeddings for graph nodes, and then uses these embeddings to predict links between two nodes. We analyze both transductive (test nodes included in the training graph) and inductive (test nodes not part of the training graph) models. We evaluate ELP when learning node embeddings using both graph connectivity and node attributes and compare to similarity-based approaches.

\section{Methods}

\subsection{Constructing  graph from the KEGG database}

The KEGG database is used to construct a data graph. Molecules in the KEGG database are represented as nodes. For each biochemical reaction, each substrate-product pair within the reaction is modeled  as an edge in the graph. As most reactions within the KEGG database are reversible, we assume that each biochemical reaction is reversible and construct a non-directional graph. Biochemical networks have  cofactor molecules (e.g., NADP, \ch{H2O}) that participate in many reactions, forming high-connectivity hub nodes within the graph \citep{ravasz2002hierarchical}.  As we aim to predict connectivity between non-cofactor metabolites, we  exclude such high-connectivity nodes and their edges from the graph.

Nodes are assigned molecular fingerprints as attributes. The fingerprints are encoded as binary vectors of fixed length $K$. We select two fingerprints that reflect the presence or absence of pre-defined structural molecular fragments: the MACCS fingerprint with $K=166$ structural keys \citep{durant2002reoptimization}, and the PubChem fingerprint with $K=881$ structural keys \citep{kim2015pubchem}. 

Enzymatic reaction data is assigned as edge attributes. Each edge is  assigned the enzyme commission (EC) number that catalyzes the associated chemical reaction. EC numbers are represented as four numbers separated by periods \citep{tipton2018brief}. For example L-lactate dehydrogenase is assigned EC number 1.1.1.27.  Each edge is also assigned a KEGG reaction class (RC) label. Each such label is associated with a group of reactions that share the same  localized structural change between a substrate and a product (e.g., the addition or removal of a hydroxyl group) \citep{kanehisa2015kegg}. Each RC label is given using a 5 digit label.  Although a reaction may be associated with one or more RC labels, each substrate-product pair is associated with only  one RC label.  If a reaction has no label, we assign it a  null label. Thus, each edge in the graph is associated with an EC label and a RC label. A graph $G=(V,E)$ therefore consists of a set of vertices $V$ and a set of edges $E$. Every node $i \in V$ represents a molecule and every edge $(i,j) \in E$ for some $(i,j) \in V$ represents an enzymatic reaction connecting two molecules $i$ and $j$ .

\begin{figure}[!t]%
\centerline{\includegraphics[width=.80\linewidth]{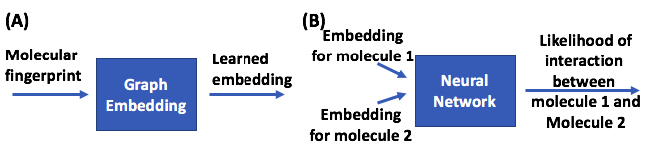}}
\caption{Overview of ELP. (A) Embeddings are learned for molecules using graph embedding.   EP computes real-value embeddings, with $d=128$.  (B) Learned embeddings are used to predict enzymatic  links.}
  \label{fig:overview}
\end{figure}

\subsection{ The ELP method}

ELP has two steps (Figure~\ref{fig:overview}): (A) learning embedding vectors of graph nodes, and (B) predicting interaction between a pair of nodes from their embedding vectors. For the first step, we use the Embedding Propagation (EP) algorithm \citep{duran2017learning}.   EP was selected because it almost consistently outperformed several other methods in the presence of node attributes on several datasets. Further, EP has the advantage of fewer parameters and hyperparameters when compared to other link prediction methods (e.g.,  \citep{perozzi2014deepwalk,tang2015line,grover2016node2vec}).  For the second step, we train a neural network that takes pairs of learned embedding vectors as input and predicts the connectivity of two molecules via a reaction.

\subsubsection{Connectivity-based learned embeddings}
The simplest form of EP is to learn a set of node embedding vectors $\bU=\{\bu_i \in \mathbb{R}^{d}: i \in V\}$, where $d$ is the embedding size.  Embeddings are randomly initialized prior to training.
Node embeddings are learned via an iterative process, by propagating forward (representations of nodes) and backward (gradients) messages between neighboring nodes. The iterative process repeats until a convergence threshold is reached.

Suppose $N(i)=\{j\in V: (i,j)\in E\}$ is the set of neighboring nodes of node $i$. The model aims to reconstruct the embeddings $\bu_i$ from the embeddings of $i$'s neighbors. The reconstructed node embedding   $\bu_i$  for node $i$ is: 
\begin{align}
\tilde{\bu}_i = \frac{1}{|\mathcal N(i)|}\sum_{j \in \mathcal N(i)} \bu_j
\end{align}

The learning objective of EP is to maximize the similarly between  $\tilde{\bu}_i$ and  $\bu_i$.  Instead of maximizing the absolute values of inner products for all such nodes, EP maximizes their values in a relative sense: the reconstruction should be more similar to the corresponding embedding vector than any other embedding vectors. The  error in reconstruction is therefore minimized through a margin-based ranking loss \citep{duran2017learning}:
\begin{align}\label{eq:obj}
\mathcal{L}=\sum_{i \in V} \sum_{j\in V, j\neq i}\max\{\gamma - \tilde{\bu}_i^\top \bu_i + \tilde{\bu}_i^\top \bu_j, 0\},
\end{align}
where $\gamma>0$ is a chosen margin hyperparameter. The objective is optimized by stochastic gradient descent. However, summing over all nodes as indicated by the inner sum is very expensive. For performance, we randomly select one node as the negative example for each real node in every iteration to compute an estimation of $\mathcal{L}$ and its gradient, as was done in \citep{duran2017learning}.

\subsubsection{Attribute-based learned embeddings}
To incorporate information from edge attributes, EP learns embedding vectors $\bZ= \{\bz_c \in \mathbb{R}^{d}: c=1,\ldots, C \}$ for the $C$ reaction labels.  The reconstructed node embedding   $\bu_i$  for node $i$ is modified as follows:
\begin{align}
\tilde{\bu}_i = \frac{1}{|\mathcal N(i)|}\sum_{j \in \mathcal N(i)} \bu_j + \alpha \:  \bz_{r(i,j)},
\end{align}
where $r_{i,k}$ is the edge label of the edge $(i,j)$ and $\bz_r(i,j)$ is the corresponding edge embedding. The hyperparameter $\alpha \in \{0, 1\}$ weights the importance of edge features. The vector $\bz_{0}$ corresponding to null edge attributes is fixed to zero  to avoid affecting the reconstruction.  Embeddings based on edge-attributes can be learned simultaneously while learning connectivity-based embeddings. While the edge embeddings are used during training, they are not used to compute the final embeddings of nodes after EP training.

EP can also  learn $K$ fingerprint embedding vectors $\bV=\{\bv_i \in \mathbb{R}^{d}: k =1, \ldots, K\}$. Specifically, the node-attribute-based embedding $\bu_i$ of a node $i$ is the mean of fingerprint embeddings $\bv_k$ corresponding to positive fingerprint entries in the fingerprint vector $\boldf_i\in \{0, 1\}^{K}$:
\begin{align}
    \bu_i^{fp} = \frac{1}{\sum_{k=1}^{K}f_{ik}} \sum_{k=1}^K f_{ik} \bv_k
\end{align}
When computing embeddings based on node-attributes we optimize the fingerprint embeddings $\bV$, instead of $\bU$, through the learning objective in Eq. \eqref{eq:obj}.   An advantage of the EP algorithm is its ability to learn only one of the node embedding types or all. If both node-attributes and connectivity embeddings are trained, we simply concatenate $\bu_i$ and $\bu_i^{fp}$ to form the final node embedding vector of node $i$ before applying the link prediction model.   L2 regularizations is applied to all variables $\bU$, $\bV$ and $\bZ$.

\subsubsection{Link prediction}
The trained node embeddings are used as inputs to a logistics link prediction model. Pairs of embeddings of nodes involved in a known reaction are positive examples; pairs of embeddings of nodes that have no or unknown interaction are treated as negative examples. To make link predictions, the neural network outputs the likelihood of an edge for every pair of input node embeddings. The final result of the model is evaluated based on the Area Under Curve (AUC) metric, wherein the false positive rate and true positive rate are evaluated at every threshold to compute the area under the Receiver Operating Characteristic (ROC) curve.

\subsection{Training and testing}
We explore two learning scenarios: transductive and inductive. In the transductive setting, the model is trained on all available nodes and evaluated on edge recovery  for a set of test edges that were withheld from training. Hence, the graph is split into training and testing sets by partitioning on the edges. To ensure that the training graph is connected,  the largest connected component is considered as the  training graph.   Therefore, the embeddings of all nodes incident to test edges are trained through the EP algorithm. ELP predicts the likelihood of enzymatic interactions between two nodes from their learned embeddings. In the inductive scenario, the model must predict interactions for one or more  {\it out-of-sample} nodes excluded from the training set. ELP therefore computes embeddings for out-of-sample nodes from their attributes and predicts possible enzymatic reactions for them.  Due to the lack of prior connectivity information for out-of-sample nodes,  only embeddings based on node-attributes are learned during training. To generate the training and testing sets, we reserve a certain portion of nodes and their incident edges for the test graph. All other nodes and edges are included in the training graph.

\begin{table}[!t]
\caption{Training and testing graph statistics for the KEGG dataset under transductive and inductive learning scenario\label{table:graphstats}} 
\resizebox{\columnwidth}{!}{
{\begin{tabular}{@{}c| cc | cc@{}}
\toprule 
 & \multicolumn{2}{c}{Training} & \multicolumn{2}{c}{Testing} \\
    & Nodes & Edges & Nodes & Edges \\
   \midrule

   \multicolumn{5}{c}{\textit{Transductive Learning}}\\
   Test Ratio 0.1 & 6,833 & 12,350 & 6,833 & 1,311\\
   Test Ratio 0.3 & 5,766 & 9,255 & 5,766 & 2,819 \\
  Test Ratio 0.5 & 4,352 & 6,136 & 4,352 & 3,552\\
    \midrule
   \multicolumn{5}{c}{\textit{Inductive Learning}}\\
   Out-of-Sample Ratio 0.05 & 7,377 & 12,867 & 387& 1,446\\
		\bottomrule
\end{tabular}}{}
}
\end{table}

\section{Results}
Once cofactors were excluded, our dataset representing the biochemical network underlying the KEGG database consisted of 7850 nodes and 14313 edges (Table~\ref{table:graphstats}).
For transductive learning experiments, we evaluate link prediction on test ratios of 0.1, 0.3 and 0.5.  In each case, the training graph represents the largest connected component after edge partitioning based on the test ratio. Nodes and edges not present in the largest connected component are discarded.  For inductive learning experiments, the ratio of out-of-sample test nodes is fixed to 0.05.  For all experiments, the embedding dimension is set to 128.  The learning rate for the EP framework is set to 0.5 and the regularization to 0.0001. The embedding vectors are trained for 100 epochs on a batch size of 512.  The NN that predicts the connectivity of two molecules based on their embeddings  consists of two hidden layers of sizes 32 and 16.  This NN is trained for 40 epochs on a batch size of 2048 with learning rate 0.2. 
The margin hyperparameter $\gamma>$ is set to 10. In  experiments using edge features, $\alpha$ is set to 1. 

\begin{table}[!t]
\caption{Link prediction AUCs for transductive learning. \label{table:transresults}}
\resizebox{\columnwidth}{!}{
 {\begin{tabular}{@{}lllllll@{}}\toprule 
 
\multicolumn{4}{c}{Model} & \multicolumn{3}{c}{AUC}\\
    \midrule
    \multirow{2}{*}{Method} & Connectivity & Node & Edge & \multicolumn{3}{c}{Test Ratios}\\
    & Embedding & Attribute & Attribute & 0.1 & 0.3 & 0.5\\
    \midrule
    \multicolumn{7}{c}{\textit{A. Connectivity-based embeddings only}}\\
    ELP & Yes & -- & -- & 0.801 & \textbf{0.789} & 0.761\\
    node2vec & Yes & -- & -- & 0.824 & 0.736 & \textbf{0.776}\\
    DeepWalk & Yes & -- & -- & \textbf{0.847} & 0.763 &  0.749\\
    \midrule
    \multicolumn{7}{c}{\textit{B. Connectivity and one additional attribute}}\\
    ELP & Yes & MACCS & -- & \textbf{0.953*} & \textbf{0.935*} & \textbf{0.900} \\
    ELP & Yes & PubChem & -- & 0.891 & 0.882 & 0.864\\
    ELP & Yes & -- & EC & 0.795 & 0.808 & 0.810 \\
    ELP & Yes & -- & RC & 0.810 & 0.798 & 0.810 \\
    \midrule
    \multicolumn{7}{c}{\textit{C. Connectivity with one node and one edge attribute}}\\
    ELP & Yes & MACCS & EC & 0.941 & \textbf{0.933} & \textbf{0.922*} \\
    ELP & Yes & MACCS & RC & \textbf{0.942} & 0.929 & 0.895 \\
    ELP & Yes & PubChem & EC & 0.892 & 0.879 & 0.867\\
    ELP & Yes & PubChem & RC & 0.892 & 0.876 & 0.859 \\
    \midrule
    \multicolumn{7}{c}{\textit{D. Embedding based on MACCS fingerprints  }}\\
    ELP & No & MACCS & -- & 0.931 & 0.916 & 0.898 \\
    ELP & No& MACCS & EC & \textbf{0.940} & \textbf{0.925} & \textbf{0.913} \\
    ELP & No & MACCS & RC & 0.939 & 0.904 & 0.896 \\
    \midrule
    \multicolumn{7}{c}{\textit{E. Embeddings based on PubChem fingerprints }}\\
    ELP & No & PubChem & -- & 0.665 & \textbf{0.709} & 0.682 \\
    ELP & No & PubChem & EC & \textbf{0.745} & 0.707 & \textbf{0.728} \\
    ELP & No & PubChem & RC & 0.728 & 0.706 & 0.720 \\
    \midrule
    \multicolumn{7}{c}{\textit{F. Jaccard index similarity scoring; no embeddings}}\\
    Jaccard & No & MACCS & -- & \textbf{0.808} & \textbf{0.778} & \textbf{0.767}\\
    Jaccard & No & PubChem & -- & 0.542 & 0.526 & 0.535\\
\bottomrule
\end{tabular}
}
}
{\small Results are partitioned to facilitate comparisons. (A) Using node embeddings representing connectivity. (B) Using connectivity embedding and one additional attribute in the form of a node or edge attribute. (C) Using node embeddings representing connectivity and one node and one edge attribute. (D) Using MACCS fingerprints and zero or more additional edge attribute with no connectivity-based node embeddings.  (E) Same as (D) but using PubChem Fingerprints. (F) Using Jaccard similarity scoring on molecular fingerprints.  Bold values in each partition indicates the best result in that train-test split. Bold values with $^*$ indicate the best overall result. The Connectivity Embedding column refers to the use of connectivity-based node embeddings.}

\end{table}

\subsection{Evaluation of Transductive Scenarios}

\begin{figure}[!t]
\centering
\begin{subfigure}
    \centering
  \includegraphics[width=.95\linewidth]{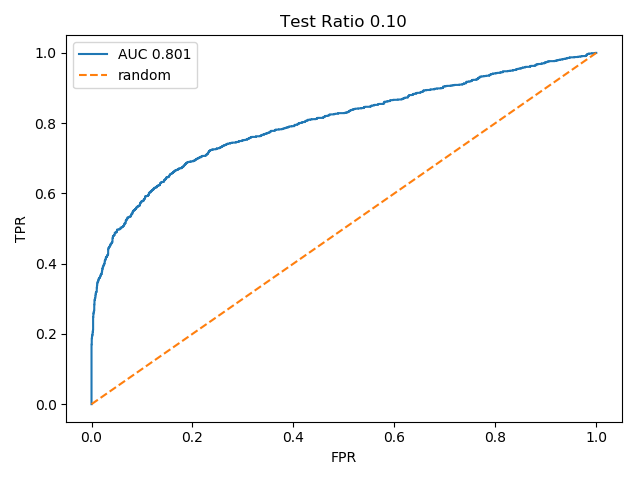}
\caption{}
  \label{fig:sub1}
\end{subfigure}%
\begin{subfigure}
\centering
  \includegraphics[width=0.95\linewidth]{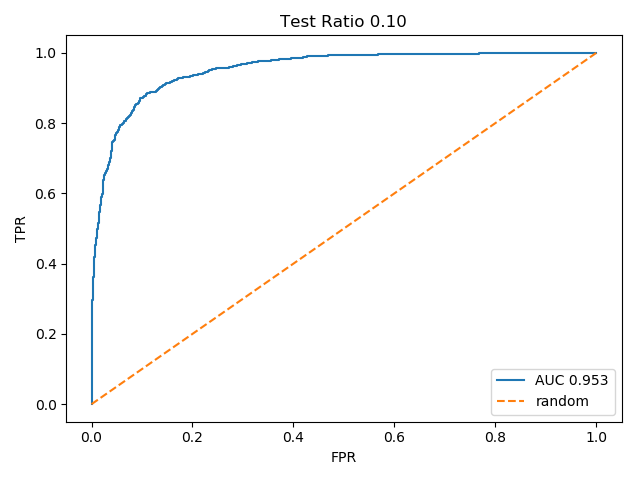}
\caption{}
  \label{fig:sub2}
 \end{subfigure}%
\caption{AUC plots for transductive learning using test ratio of 0.1. (A) graph connectivity only, and (B) graph connectivity with MACCS fingerprint as node attributes. }\label{fig:AUC}
\end{figure}

Results for several transductive are reported (Table \ref{table:transresults}, partitions (A)-(E)).  
For almost all cases, a smaller test ratio, results in higher AUC. This result is expected as a larger training graph better informs prediction. 
When performing connectivity-based prediction (partition A), ELP, node2vec \citep{grover2016node2vec}, and DeepWalk  \citep{perozzi2014deepwalk} performed comparably, with less than < 05. AUC differences for each  test ratio.                                 
Per partition (B), concatenating the MACCS fingerprints embeddings to those based on connectivity improves the AUC to 0.953 from 0.801 for test ratio 0.1. Gains of more than 0.1 are recorded for the other test ratios. Not the PubChem fingerprint, nor any of the edge attributes enhance the AUC as much as the MACCs fingerprint.  Further, there is little difference between using the two edge embeddings, and both seem to add little enhancement to the AUC results over the connectivity-only ELP. Per partition (C), adding an edge and a node attribute simultaneously to graph connectivity results in a slight decrease in performance when compared to using connectivity and MACCS fingerprints as node attributes for test ratio 0.1, but adding EC edge attributes improves performance for test ratio 0.5.   Per partition (D), using embeddings for node attributes only provides results comparable to that of using graph connectivity and the MACCS fingerprint for all test ratios (e.g., 0.931 vs 0.953 for 0.1 test ratio).  Comparing partitions (D) and (E), link prediction using the MACCS fingerprint outperforms the scenario when using PubChem fingerprints. When using embeddings for edge attributes, more pronounced AUC improvements are observed for embeddings based on the PubChem fingerprint vs the MACCS fingerprint.  
 When using the Jaccard index to compute substrate-product similarity for each link, as in partition (E), the AUC is 0.808 for the MACCS fingerprints, while significantly lower when using the PubChem fingerprint.    Figure~\ref{fig:AUC} presents  plots for two scenarios using ELP: (A) connectivity only and (B) connectivity with MACCS fingerprints as node attributes. The former plot reveals that the lower AUC performance is mostly attributed to having a higher FPR when there is a higher TPR. In other words, we can achieve an almost .50 TPR at little cost (little sacrifice in FPR), but as the need to observe improvement in TPR increases, the FPR rises dramatically. 

\subsection{Evaluation of Inductive Scenarios}

Several inductive scenarios were investigated (Table \ref{table:indresults}). In these scenarios, 5\% of all nodes were removed from the graph during training. 
ELP based on MACCS node attributes achieves an AUC of 0.921. When compared to the similar transductive scenario, ELP based on MACCS node attributes, the AUC for the inductive scenario is lower than the AUC for the 0.1 and 0.3 test ratios (AUC of .953, 0.935, respectively), but higher than the AUC for the 0.5 test ratio (AUC of 0.900).  Using the PubChem fingerprint results in a lower AUC. 
Similarity analysis based on the Jaccard index achieves an AUC of 0.744 when using MACCS fingerprints and results in a much lower AUC when using the PubChem fingerprints. This analysis indicates that more informative embeddings even for out of sample nodes are best predicted through ELP.

\begin{table}[!t]
\caption{Link prediction AUCs for inductive learning. \label{table:indresults}}
\resizebox{\columnwidth}{!}{
 {\begin{tabular}{@{}cccc@{}}\toprule 
    Method  & Connectivity Embedding & Node  Attribute& AUC \\
    \midrule
        \multicolumn{4}{c}{\textit{A. Embeddings based on node attributes}}\\
    ELP& Yes & MACCS&\textbf{0.921}\\
    ELP & Yes& PubChem &0.605 \\
        \midrule
            \multicolumn{4}{c}{\textit{B. Jaccard index similarity scoring }}\\
    Jaccard & No & MACCS&0.744\\
    Jaccard & No & PubChem & 0.553\\
    \bottomrule
\end{tabular}}
}
{The best AUC is denoted in bold.}

\end{table}

\subsection{Additional application: graph embedding applied to biochemical network visualization }
\begin{figure}[!t]%
\centering
 \begin{subfigure}
  \centering
  \includegraphics[width=.8\linewidth]{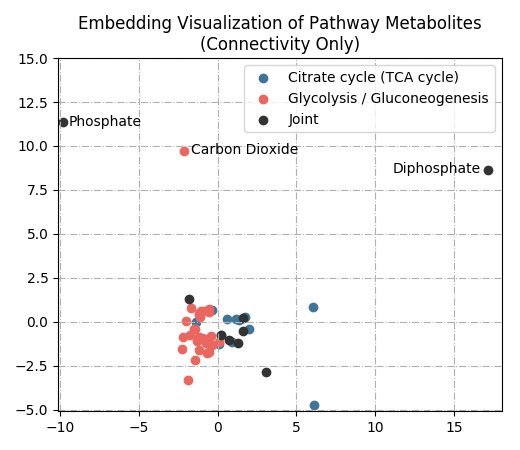}
\caption{}
  \label{fig:vsub1}
\end{subfigure}
  \begin{subfigure}
  \centering
  \includegraphics[width=.8\linewidth]{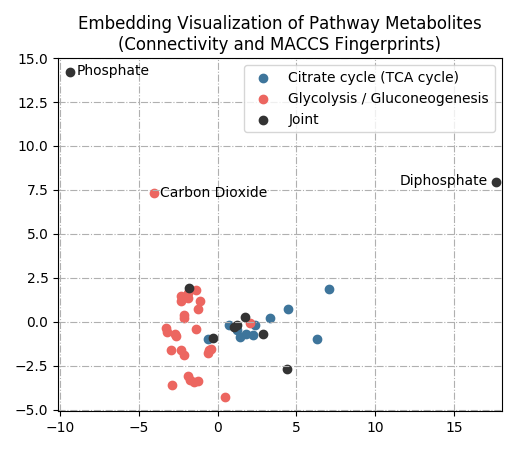}
\caption{}
  \label{fig:vsub2}
\end{subfigure}

\caption{2D-embedding plot for two transductive scenarios using ELP: (A) using graph connectivity only, and (B) using both graph connectivity and MACCS fingerprints.  }\label{fig:viz}
\end{figure}

To  further illustrate the importance of graph embedding in the context of biochemical networks, we show how graph embedding can be used for visualization. Figure~\ref{fig:viz} presents a visualization of the embeddings for two reference pathways, the citrate cycle (TCA) cycle, and Glycolysis/Gluconeogenesis, as documented in the KEGG database.  The resulting subgraph for the TCA cycle consists of 21 nodes and 44 edges, while the subgraph for Glycolysis/Gluconeogenesis pathway consists  37 nodes and 96 edges.  Nine compounds are common to both pathways, and  include  phosphate, diphosphate, pyruvate, thiamine diphosphate, lysine, oxaloacetate, and phosphoenolpyruvate.  These compounds contribute to 14 edges that overlap in both pathways. To visualize  embeddings of these metabolites, we reduce the dimensionality of the embeddings to 2 via Principal Component Analysis (PCA). For the connectivity only plot (Figure~\ref{fig:viz}A), we observe tight clustering of metabolites within each pathway, while we observe looser clustering when using MACCS fingerprints as node attributes (Figure~\ref{fig:viz} B).  Nodes that are embedded far away from the clusters, phosphate, diphosphate, and carbon dioxide, exhibit high connectivity within the KEGG graph, with  node degrees of 408, 320, and 494, respectively.  On the contrary, other nodes within the KEGG graph have an average degree of 27, with most nodes having degrees under 30.  

\section{Conclusion}

This work uses embedding propagation to learn molecular representations that capture both graph connectivity, enzymatic properties, and structural molecular properties.  
We show that link prediction using only graph connectivity is on par with using molecular similarity. Additionally, we show high accuracy in link prediction when using both graph connectivity and molecular attributes. This work has broader and  practical impact. ELP  can be used to guide many biological discoveries and engineering applications such as identifying catalyzing enzymes when constructing novel synthesis pathways or predicting interaction between microbes and human hosts. Graph embedding can be used for other applications such as biochemical network visualization, as demonstrated herein, and identifying synthesis routes for synthetic biology. Further, while our approach is applied to biochemical enzymatic networks, it can enhance link prediction in chemical networks, where rule-based and path-based link prediction respectively yielded 52.7\% and 67.5\% prediction accuracy \citep{segler2017modelling}.

\section*{Funding}
This research is supported by NSF,  Award  CCF-1909536, and also by NIGMS of the National Institutes of Health, Award  R01GM132391. The content is solely the responsibility of the authors and does not necessarily represent the official views of the National Institutes of Health.

\printbibliography[title={Bibliography}]

\end{document}